\documentclass[12pt]{article}
\usepackage{amssymb}
\usepackage{amsmath}
\usepackage{amsfonts}
\usepackage{graphicx}

\def\vbs{\vspace{1cm}}
\def\bs{\vspace{0.5cm}}

\def\beq{\begin{equation}}
\def\eeq{\end{equation}}
\def\beqa{\begin{eqnarray}} 
\def\eeqa{\end{eqnarray}}
\def\lsim{\mathrel{\rlap{\lower4pt\hbox{\hskip1pt$\sim$}}
    \raise1pt\hbox{$<$}}}
\def\gsim{\mathrel{\rlap{\lower4pt\hbox{\hskip1pt$\sim$}}
    \raise1pt\hbox{$>$}}}
\begin{document}

\begin{center}

{\Large \bf WMAP Bounds on Braneworld Tachyonic Inflation}
\vbs

M. C. Bento\footnote{Also at CFIF, Instituto Superior T\'ecnico, Lisboa.
  Email address: bento@sirius.ist.utl.pt},
N. M. C. Santos 
\footnote{Also at CFIF, Instituto Superior T\'ecnico, Lisboa.
 Email address: ncsantos@cfif.ist.utl.pt }
and A. A. Sen
\footnote{Also at CENTRA, Instituto Superior T\'ecnico, Lisboa.
 Email address: anjan@x9.ist.utl.pt}
\vspace{0.5cm}

{ Departamento de F\'\i sica, Instituto Superior T\'ecnico \\
Av. Rovisco Pais 1, 1049-001 Lisboa, Portugal}

\bs
{\bf Abstract}

\end{center}
\noindent
We analyse the implications of the Wilkinson Microwave Anisotropy
Probe (WMAP) results for a braneworld tachyonic model of inflation.
We find that WMAP bounds on $n_s$ allow us to constrain significantly 
the parameter space of the model; in particular, extremely weak string
coupling is required,  $g_s \sim 10^{-15}$. Moreover, our analysis 
shows that the running of the scalar spectral index is within the
bounds determined by WMAP for the allowed range of model parameters;
however, it is not possible to obtain $n_s>1$ on large scales and 
$n_s<1$ on small scales.

\vspace{1cm}
PACS number(s): 98.80.Cq

\section{Introduction}

Recent measurements of the cosmic microwave background anisotropies
lend powerful support to the inflationary paradigm {\it i.e.}  the
existence of an epoch of accelerated expansion in the very early
universe which dynamically solves the cosmological puzzles such as the
homogeneity, isotropy and flatness of the universe \cite{Guth:1980zm}.
During this accelerated expansion phase, primordial quantum
fluctuations of fields are amplified and act essentially as seeds for
structure formation in the universe.  In particular, the remarkably
accurate data set obtained by the WMAP satellite has made it possible
to significantly constrain inflationary models, on the basis of their
predictions for the primordial power spectrum of density perturbations
\cite{Bennett:2003bz,Hinshaw:2003ex,Spergel:2003cb}.  WMAP data
provides no indication of any significant deviations from gaussianity
and adiabacity; moreover, it allows for very accurate constraints on
the spectral index, $n_s$, and its running, $\alpha_s$
\cite{Peiris:2003ff} \beq n_s = 1.10^{+0.07}_{-0.06}~,\quad
\alpha_s\equiv dn_{s}/d \ln k =-0.042^{+0.021}_{-0.020}~, \eeq on the
scale $k_{0} = 0.002$ Mpc$^{-1}$. The data suggest but do not require
that, at $2\sigma$ level, $n_s$ runs from $n_s>1$ on large scales to
$n_s<1$ on small scales; we should stress, however, that the
statistical significance of this result is not entirely clear, as
pointed out in
Refs.~\cite{Seljak:2003jg,Mukherjee:2003ag,Barger:2003ym}.

Recently, following the pioneering work by A. Sen in understanding the
role of the tachyon condensates in string theory \cite{Sen:2002},
there has been considerable interest in developing models of inflation
driven by such a field \cite{Tachyonall,Kofman:2002rh}.  One of the
main problems in modeling tachyon inflation in standard Einstein
gravity is that one cannot obtain sufficient inflation to solve the
cosmological problems for a reasonable choice of model parameters
\cite{Kofman:2002rh}. Moreover, in this scenario, inflation occurs at
super-Planckian values of the brane energy density, making the
effective four-dimensional gravity theory unreliable near the top of
the potential.

In a recent work \cite{Bento:2002np}, an alternative framework for
 inflation driven by the tachyon field was proposed, where the tachyon
 is seen as a degree of freedom on the visible three dimensional
 brane.  For this purpose, a specific braneworld scenario is
 considered, the Randall-Sundrum Type II model (RSII), which implies that the
 dynamics on the brane is described by a modified version of the
 Einstein equations
 \cite{Shiromizu:1999wj,Binetruy:1999hy,Flanagan:1999cu}.  In this
 context, it has been shown that one can have a successful
 inflationary scenario, where there is sufficient inflation while the
 energy density remains sub-Planckian.

In this article, we study the implications of WMAP results for the
model 
proposed in 
Ref.~\cite{Bento:2002np}.

\section{Braneworld Tachyonic Inflation}

Unstable non-BPS D-branes are characterized by having a single tachyon 
mode living on their world volume. The effective field theory action for this
 tachyon field $T$ on a D3 brane, computed with
 the bosonic string field theory, 
around the top of the potential, is given by 
\cite{Gerasimov:2000zp,Kutasov:2000qp}

\beq
S_B = \tau_3 \int d^4 x\exp(-T)[l_s^2 \partial_{\mu} T \partial^{\mu} T
+ (1 + T)]~,
\label{eq:action}
\eeq
where $\tau_3$ is the tension of the (unstable) D-brane and is given by

\beq
\tau_3={{M_s^4}\over{(2\pi)^3 g_s}}~,
\label{eq:tau3}
\eeq
 $g_s$ being  the string coupling and $M_s$ the  string mass scale, given by 
$M_s=l_s^{-1}$, where $l_s$ is the fundamental string length scale.
On the other hand, the 4D Planck mass is obtained via dimensional
 reduction, leading to  \cite{Jones}

\beq
M_P^2=8 \pi {M_s^2  v\over g_s^2}~,
\label{eq:Mp2}
\eeq 
where $ v=(M_s r)^d/\pi$ is a dimensionless parameter corresponding
to the volume of the 22D space (as we are considering here
the bosonic string case) transverse to the brane, $r$ is the radius of this
compactified volume and $d$ is the number of compactified dimensions.
 Usually one assumes that $r\gg l_s$,
i.e. ${ v }\gg 1$, in order to be able to use the 4D effective theory
\cite{Jones}.

One can also write down the closed form expression for the above action
 including all the higher powers of $\partial_{\mu} T$ in a
 Born-Infeld form \cite{Sen:2002}. However, we are interested in the
 early time evolution of the tachyon field, when it is slightly
 displaced from the top of the potential, where the time derivatives of
 the tachyon field turn out to be small, and therefore we can safely take the
 simpler action (\ref{eq:action}) for all practical purposes.

Notice that the kinetic term has a nonstandard form due to the factor 
$\exp(-T)$; it is, however, possible to write this term in  canonical
form  via a field  redefinition

\beq
\phi = \exp(-T/2)~.
\label{eq:red}
\eeq
As a consequence, the potential becomes

\beq
V(\phi) = -\tau_{3}\phi^{2}\ln(\phi^2/e)~.
\label{eq:newv}
\eeq
Notice that $0<\phi<1$; the limit $\phi\rightarrow 0$ ($T \rightarrow
 \infty$) corresponds to the stable vacuum to which the tachyon condensates.

Hereafter, we shall consider the tachyon field, with potential given
 by
 Eq.~(\ref{eq:newv}), as the inflaton.
We also assume that our universe is a 3D hypersurface within
a 5D spacetime, in which the bulk contains a negative
cosmological constant (anti-de Sitter bulk); moreover, we assume that
the matter fields are confined to our 4D universe. In this
case, the Friedmann equation in 4D  acquires an
extra term, becoming
\cite{Shiromizu:1999wj,Binetruy:1999hy,Flanagan:1999cu}

\beq
H^2 =  \left({8 \pi \over 3 M_P^2}\right) \rho
+ \left({4 \pi \over 3 M_5^3}\right)^2 \rho^2 +{\Lambda \over 3} + 
{\epsilon \over a^4}~,
\label{eq:H2}
\eeq
where $\Lambda$ is the 4D effective cosmological constant, which is
related to the 5D cosmological constant  and 
the brane tension, $\lambda$, through

\beq
\Lambda = {4 \pi \over M_5^3} \left(\Lambda_5 + {4 \pi \over 3 M_5^3}~
\lambda^2 \right)~.
\label{eq:Lam}
\eeq
The brane tension relates the  Planck mass in 4D and 5D via

\beq
M_P = \sqrt{{3 \over 4 \pi}} {M_5^3 \over \sqrt{\lambda}}~.
\label{eq:MP}
\eeq

Assuming that the 4D cosmological constant cancels out via some
mechanism, the last term in Eq.~(\ref{eq:H2}), which represents the
influence of bulk gravitons on the brane, rapidly becomes unimportant
after inflation sets in. In this
 case, the Friedmann equation becomes 

\beq
H^2 = {8 \pi \over 3 M_P^2} \rho \left[1 + {\rho \over 2 \lambda}\right]~.
\label{eq:H2new}
\eeq

The new term in $\rho^2$ is dominant at high energies, but quickly
 decays at lower energies, and the usual 4D FRW cosmology is recovered.
Since the scalar field is confined to the brane, its field equation has
 the standard form

\beq
\ddot \phi + 3 H \dot \phi + {1\over{8\tau_3 {l_s}^2}}{d V\over d \phi} = 0~.
\label{eq:eqmphi}
\eeq
except for the the extra factor, ${1/{8\tau_{3}l_{s}^{2}}}$, 
appears due to the field redefinition of Eq.~(\ref{eq:red}).
We will consider the slow-roll approximation, in which case the inflationary
parameters can be written as \cite{Maartens:1999hf}

\beq
\epsilon \equiv {M_{P}^2 \over 16\pi}{1\over{8\tau_{3}l_{s}^{2}}}
\left( {V' \over V}\right)^2  {1+{V/ \lambda}\over(1+{V/ 2 \lambda})^2}~,
\label{eq:epsilon}
\eeq

\beq
\eta \equiv {M_{P}^2\over{8\pi}}{1\over{8\tau_{3}l_{s}^{2}}}
{V'' \over V}   {1 \over 1+{V/ 2 \lambda}}~,
\label{eq:eta}
\eeq

\beq
\xi \equiv{M_{P}^4\over{(8\pi)^2}}{1\over{\left(8\tau_{3}l_{s}^{2}\right)^2}}
{V' V''' \over V^2}   {1 \over \left(1+{V/ 2 \lambda}\right)^2}~,
\label{eq:xi}
\eeq
where prime indicates a $\phi$-derivative. 

The number of e-folds during inflation is given by 
$N =\int_{t}^{t_{\rm f}} H d\bar{t}$, which,
 in the slow-roll approximation, becomes \cite{Maartens:1999hf}

\beq
\label{eq:N}
N(\phi) \simeq - (8\tau_{3}l_{s}^{2}){8\pi  \over M_{P}^2}
\int_{\phi}^{\phi_{\rm f}}{V\over V'}
\left[ 1+{V \over 2\lambda}\right]  d\bar{\phi}~,
\eeq
 where $\phi_f$ is  the value 
of $\phi$ at the end of inflation, which  can  be obtained from the condition
\beq
{\rm max}\{\epsilon(\phi_f),|\eta(\phi_f)|\}= 1~.
\label{eq:phif}
\eeq
The amplitude of scalar  perturbations is  given 
by \cite{Maartens:1999hf}

\beq
A_s^2 = \left . 8\tau_{3}l_{s}^{2}{512\pi\over 75 M_P^6}{V^3\over V^{\prime2}}
\left[ 1 + {V \over 2\lambda} \right]^3 \right |_{k=aH}~,
\label{eq:As}
\eeq
where the subscript $k=aH$ means that the amplitudes should be evaluated at
Hubble radius crossing. The amplitude of tensor perturbations
is given by \cite{Maartens:1999hf}

\beq
A_t^2 =\left . {32\over 75 M_P^4} V
\left[ 1 + {V \over 2\lambda} \right] F^2\right |_{k=aH}  ~,
\label{eq:At}
\eeq
where

\beq
F^{-2}=\sqrt{1+s^2} -s^2 \sinh^{-1}\left({1\over s}\right)
\eeq
and
\beq
s\equiv \left( {3 H^2 M_P^2\over 4 \pi \lambda}\right)^{1/2}~.
\eeq
In the low energy limit ($s\ll 1$), $F^2\approx 1$, whereas
$F^2\approx 3 V/2\lambda$
 in the high energy limit. We parametrize the tensor power spectrum
 amplitude by the tensor/scalar ratio

\beq
r_s\equiv 16 {A_t^2\over A_s^2}~,
\eeq
where we have chosen the normalization  so as to
be consistent with the one of Ref.~\cite{Peiris:2003ff}, in the
low-energy
 limit.

The scale dependence of the scalar perturbations is described by the spectral 
tilt

\beq
n_s-1 \equiv {{d \ln A_s^2}\over{d \ln k}}=-6 \epsilon + 2 \eta~,
\label{eq:ns}
\eeq
and the running of the spectral index can be written as

\beq
\alpha_s={{d n_s}\over{d \ln k}}=16 \epsilon \eta - 18 \epsilon^2-2 \xi~.
\label{eq:alphas}
\eeq

\section{Constraints from Inflationary Observables}

 The slow-roll parameters $\epsilon$
and $\eta$,
 in this model, are  given by

\beq
\epsilon (\phi) =-{2\pi^3 v\over g_s}
{(1+\ln(\phi^2/e))^2\over \phi^2\ln^2(\phi/e)}{ 1-\beta
  \phi^2\ln(\phi/e)
 \over[1-\beta/2~ \phi^2\ln(\phi/e)]^2}~,
\label{eq:neweps}
\eeq

\beq
\eta (\phi) = -{2 \pi^3 v\over g_s}
{ (3+\ln(\phi^2/e))\over \phi^2\ln(\phi/e)[1-\beta/2~\phi\ln(\phi/e)]}~;
\label{eq:neweta}
\eeq
taking into account that
inflation starts near the top of the potential, where
$\phi\approx 1$, the inflationary condition $|\eta|\ll 1$ ($\epsilon
\ll 1 $ is trivially satisfied since $\epsilon \approx 0$ near to top
of the potential)
 leads to a lower bound on $g_s$, namelly

\beq
g_s\gg \frac{4 \pi^3 v} {1+{\beta\over 2}}~,
\label{eq:gs1}
\eeq
where $\beta\equiv\tau_3/\lambda$.
On the other hand, for the 4D gravity theory to be applicable, one
should have

\beq
{\tau_3\over M_p^4}={g_s^3\over 512 \pi^5 v^2} \ll 1~,
\label{eq:tau3b}
\eeq
leading to an upper bound on $g_s$

\beq
g_s^3 \ll 512 \pi^5 v^2~.
\label{eq:gs12}
\eeq
Combining Eqs.~(\ref{eq:tau3b}) and (\ref{eq:gs1}), we get

\beq
{\tau_3\over M_p^4}={\pi^4 v\over 8\left(1+{\beta\over 2}\right)^3} \ll 1~,
\label{eq:tau3c}
\eeq
which, as $v\gg 1$, requires that $\beta \gg 1$, i.e.  $V/\lambda\gg 1$,
 meaning that the high energy regime is required in order for inflation
 to take place. Accordingly, we shall use the high energy approximation
 hereafter. Notice that 
the condition $\beta \gg 1$ also
 implies that the bound of Eq.~(\ref{eq:gs1}) can be written as

\beq
g_s\gg 248 { v\over \beta }~.
\label{eq:gs2}
\eeq
 
Using the high-energy approximation, the slow-roll parameters become

\beq
\epsilon (\phi) \simeq -{8\pi^3\over\gamma}
{(1+\ln(\phi^{2}/e))^{2}\over\phi^4\ln^3(\phi^{2}/e)}~,
\label{eq:epsilon2}
\eeq

\beq
\eta (\phi) \simeq -{4 \pi^3\over \gamma}
{ (3+\ln(\phi^2/e))\over \phi^4\ln^2(\phi^2/e)}~,
\label{eq:neweta2}
\eeq
and 

\beq
\xi(\phi) \simeq -{32 \pi^6\over \gamma^2}
  {(1+\ln(\phi^2/e))\over \phi^8
\ln^4(\phi^2/e)} ~,
\label{eq:newxi}
\eeq
where $\gamma \equiv  {g_{s}\beta / v}$.

The total number of e-folds during inflation is given by

\beq
N={\gamma\over  4 \pi^3}  \int_{\phi_i}^{\phi_f} {\phi^3\log^2 \phi^2/e
\over \log (\phi/e + 1)}d \phi~;
\label{eq:Ntot}
\eeq
the condition that N  should be at least 70, the minimum amount of
inflation required to solve the horizon problem, implies
 $\gamma\geq 2818$, which  leads to  a lower bound on $g_s$ that
is stronger than the one of Eq.~(\ref{eq:gs2})

\beq
g_s\geq 2818 {v\over \beta}~.
\label{eq:gs4}
\eeq

The number of e-folds between the time the 
scales of interest leave the horizon and the end of inflation, $N_k$,
 is given by Eq.~(\ref{eq:Ntot}), with $\phi_i \rightarrow \phi_k$,
where $\phi_k$ is the value of $\phi$ at scale $k=aH$.

The amplitude of tensor perturbations can be written as

\beq
A_t^2\simeq {3\over 25}\sqrt{3\over \pi}{g_s^3 \beta\over v}
 {H_k^3\over M_s^3}~;
\label{eq:at2m}
\eeq
using the observational constraint  $A_t \lsim 10^{-5}$ together with 
$H_k\approx H_e$ (the Hubble parameter is not expected to
change significantly between horizon crossing  and the end of inflation)
and $H_e\approx M_s$
 (which derives from the breaking of the slow-roll condition  $H^2\gg 
\vert\dot H \vert$ together with  Eqs.~(\ref{eq:H2new}) and (\ref{eq:eqmphi})),
 Eq.~(\ref{eq:at2m}) leads to a
 further upper bound on $g_s$

\beq
g_s^3\lsim 7.3 \times 10^{-19} {v\over \beta}~.
\label{eq:gs3}
\eeq
Notice that, since $\beta\gg 1$, this bound is compatible with previous 
bounds, Eqs.~(\ref{eq:gs2}) and (\ref{eq:gs4}), whereas in standard
 cosmology the
 corresponding bounds are 
not compatible, which is at the heart of the well known difficulties of
 tachyonic inflation in that context \cite{Kofman:2002rh}. 

Finally, the amplitude of scalar perturbations is given by

\beq
A_s^2 (\phi_k) \simeq {\gamma^3 g_s\over 19200 \pi^8}
{\phi_k^{10}\ln^6(\phi_k^{2}/e)\over{(\ln(\phi_k^{2}/e) +1)^2}}~,
\label{eq:newAs}
\eeq
and  the ratio between the tensor and scalar amplitudes can be wriiten
as 

\beq
r_s (\phi_k) = -{192\pi^3\over \gamma}{(1+\ln(\phi_k^{2}/e))^2 \over
  {\phi_k^{4} \ln^3(\phi_k^{2}/e)}} ~.
\label{eq:newrs}
\eeq

\section{WMAP constraints}

We have studied the dependence of the scalar spectral index, its
 running and the  tensor/scalar ratio, on parameters $\gamma$ and 
$N_*=N_k(k=0.002~\mbox{Mpc}^{-1})$, where $k=0.002~\mbox{Mpc}^{-1}$ is
the  scale best probed by WMAP observations.

Notice that we have chosen to vary $N_\star$ since, as recently discussed in
Refs.~\cite{Dodelson:2003vq,Liddle:2003as}, there are considerable
uncertainties in the determination of this quantity, which depends,
for instance, on the mechanism  ending inflation and the reheating process. 
 In Ref.~\cite{Dodelson:2003vq} it is  shown that, for a vast class of
slow-roll models,  within standard cosmology, one should have
$N_\star<67$. In Ref.~\cite{Liddle:2003as}, a plausible
upper limit is found,  $N_\star\lsim 60$, with the expectation that
the actual value will be up to $10$ below this. However, the authors
stress that there are several ways in which $N_\star$ could lie
outside that range in either direction. If inflation takes place within  the
braneworld context, in the high energy regime, the expansion laws
 corresponding to matter and radiation domination are slower 
than in standard cosmology, which implies a greater change in $aH$
relative to the change in $a$,  requiring a large value of
$N_\star$. In Ref.~\cite{Wang}, the upper bound $N_\star< 75$ is found
in the context of brane-inspired cosmology.

\begin{figure}
\includegraphics[width=13cm]{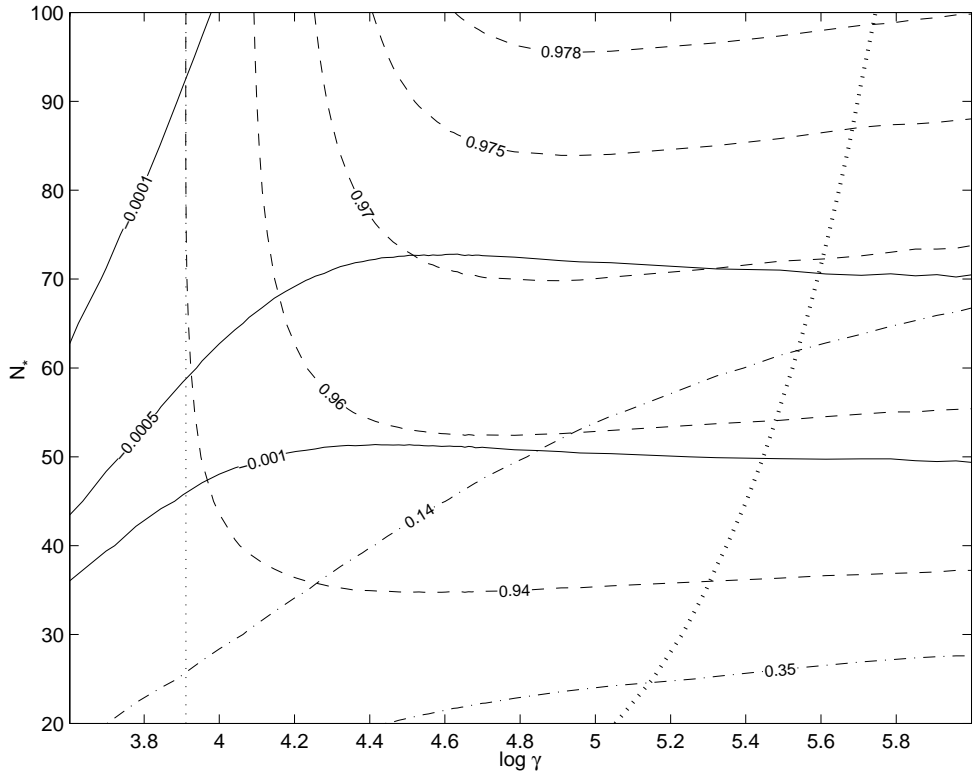}
\caption{Contour plots of $n_s$ (dashed),
$\alpha_s$ (full) and $r_s$ (dot-dashed) in the $(\gamma ,N_\star)$
  plane.  The
bold dotted curve separates the region where the model behaves as
class A ($\eta<0$, left region) from the region where it behaves as class B
($0\leq \eta<2\epsilon$, right region). The vertical dotted line
corresponds to $\gamma=8150$, see Eq.~(\ref{eq:gamma}).}
\label{fig:figure1}
\end{figure}
In Ref.~\cite{Peiris:2003ff}, inflationary slow-roll models were 
classified according to the curvature of the potential, $\eta$.
For models with $\eta<0$ (class A in \cite{Peiris:2003ff}), the bounds are
\beq
0.94 \leq n_s \leq 1.00~,~~~-0.02 \leq \alpha_s \leq 0.02~,~~~r_s \leq 0.14~,
\label{eq:WMAPConst1}
\eeq
whereas, for $0 \leq \eta \leq 2\epsilon$ (class B in \cite{Peiris:2003ff}), 

\beq
0.94 \leq n_s \leq 1.01~,~~~-0.02 \leq \alpha_s \leq 0.02~,~~~r_s \leq 0.35~.
\label{eq:WMAPConst2}
\eeq
at $95\%$
CL. The model we are studying is basically class A but, for $\phi < 1/e$,
it becomes class B. 
 
In Figure \ref{fig:figure1}, we show contours for different values of
$n_s$ (dashed), 
$\alpha_s$ (full) and $r_s$ (dot-dashed) in the $(\gamma,N_\star)$ plane.
 We also show the dividing line
(bold dotted)
between the regions where the model behaves as class A (region on the left)
and class B (region on the right).
We have checked that it is not possible to get
  $n_s$ larger than 1, even if we increase the range of
parameter $\gamma$; in fact, it is clear from Eqs.~(\ref{eq:epsilon2}),
(\ref{eq:neweta2}) and (\ref{eq:ns}) that $n_s < 1- {16 \pi^3 /\gamma}$.
 Moreover, Figure \ref{fig:figure1} shows that the constraint $n_s\leq
 0.94$ leads to  a lower bound on $\gamma$, indicated by the vertical dotted
  line, namely

\beq
 \gamma\gsim 8150~.
\label{eq:gamma}
\eeq
Regarding the observational bounds on $\alpha_s$ and $r_s$,  they
are clearly satisfied in this  model although small negative values
for the running are preferred. In particular, it is not possible 
to get $\alpha_s >0$; in fact,  even $\alpha_s=0$ can 
only be obtained in the limit $\gamma\rightarrow \infty$.

We find that, for the parameter range of Figure~\ref{fig:figure1},
 the total number of e-folds, $N$, is always larger than 70, e.g. we get
$N (\gamma= 8150)\approx 203$ (we have assumed that $\phi$ starts 
rolling very near the top of the potential, $\phi_i=0.999$).

\begin{figure}
\includegraphics[width=16cm]{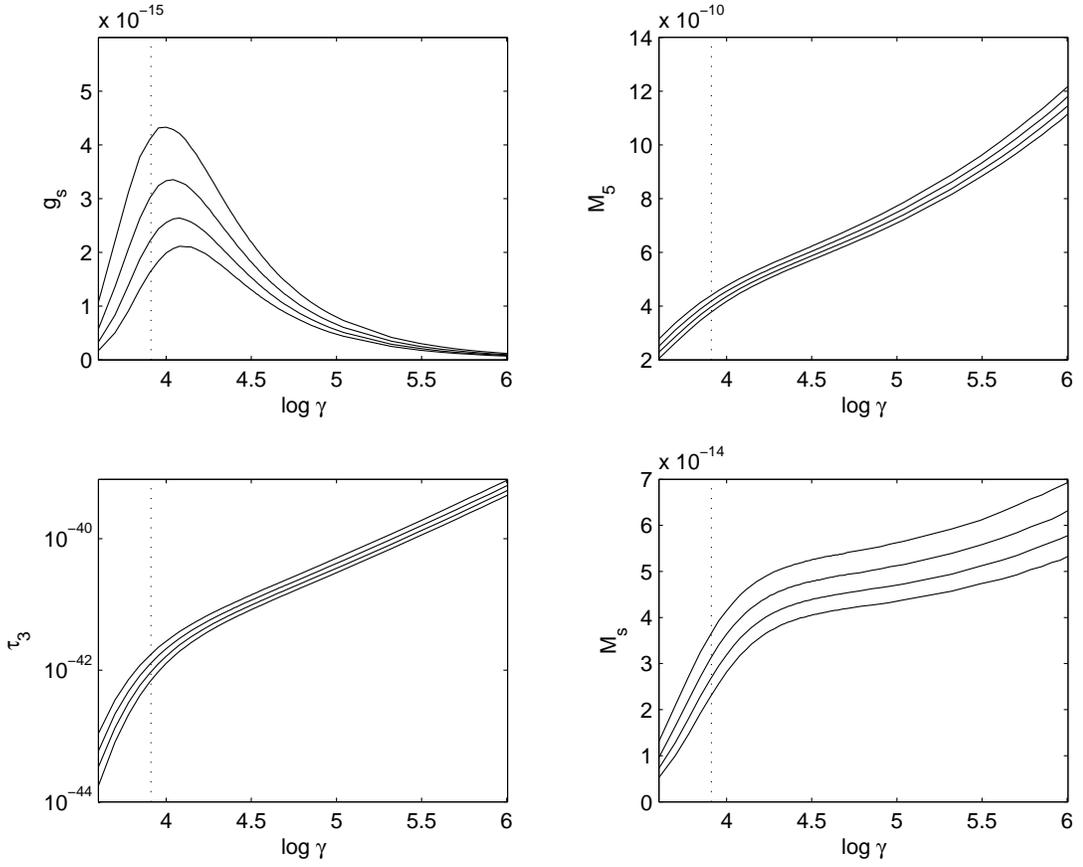}
\caption{Plots of  $g_s$ and $M_5$,  $M_s$,  $\tau_3$ (in Planck  units)
as a function of $\gamma$ for
 $N_\star=55,60,65,70$ (respectively, from top to bottom). The vertical
  line corresponds to $\gamma=8150$, see Eq.~(\ref{eq:gamma}).}
\label{fig:figure2}
\end{figure}
Combining the constraints of Eqs.~(\ref{eq:gamma}) and  (\ref{eq:gs3})
together with $v\gg 1$, we get

\beq
g_s\lsim 10^{-11}, \qquad \beta\gg 10^{15}~,
\label{eq:beta}
\eeq
from which we can, in turn, obtain
upper
 bounds on the mass scales of interest;
in Figure \ref{fig:figure2}, we have plotted the upper bounds on
 $M_5$, $M_s$ and $\tau_3$ (in
Planck units) as a function of
 $\log \gamma$, for $N_\star=55,60,65,70$. Also shown is $g_s$ as a function
 of $\log\gamma$, as derived from Eq.~(\ref{eq:newAs}) and the COBE
 normalization, $A_s(\phi_\star)\approx 2\times 10^{-5}$. 
\section{Conclusions}

We have examined the conditions for the onset of  tachyon-driven
 inflation in the context of braneworld cosmology and found that
the high-energy regime of the theory is required.

 We have also studied the 
implications of WMAP results for this model and found that
 the main constraints come from
 WMAP's
 lower bound on $n_s$, which  implies that $\gamma\gsim 8150$.
The $\alpha_s$ and $r_s$ bounds do not further 
constrain the parameter space.
Regarding  the possibility that the spectral index runs
from red on small scales to blue on large scales we conclude that,
 although $n_s$ decreases with
$k$, this is not possible since $n_s <1$. 

We have checked that the string energy density
remains sufficiently below the
 Planck scale, 
$\tau_3 \ll 10^{-42} - 10^{-40}~M_P^4$, so that the use 
of the low energy 4D gravity theory is vindicated. Also, in this
model, we get $M_5\ll  10^{-10} M_P$ and $M_s\ll 10^{-14}~M_P$. 

 We would like to stress that, in order to have  successful 
 inflation driven by the tachyon,
 in the RSII braneworld context, 
  extremely weak string coupling is required,  $g_s \sim 10^{-15}$.
Notice that we are modelling the tachyon for an  unstable D-brane in the
open string field theory but one should have closed string radiation
from the D-brane as the tachyon rolls down to its minimum. The fact
that the D-brane energy density is carried away by the closed string
seems to invalidate the open string analysis; however, it has recently been
 conjectured  \cite{sen:2003} that a full quantum open
string field theory can describe the full dynamics of an unstable
D-brane which is dual to its description in terms of closed string
emission.  Furthermore, this conjecture holds in the classical limit
which is the case we are interested in, i.e. for a very weak string coupling.

 Finally, we should mention that, if we consider the tachyon as the inflaton in
standard cosmology, due to the weak string coupling constraint,
it is not even possible for  inflation to start \cite{sen:2003}, which is also
apparent from Eq.~(\ref{eq:gs1}) with $\beta=0$ (the standard
cosmology result). We have shown
that, in the RSII braneworld scenario, it is possible to achieve
  successful inflation in a weak coupling regime.

\vbs

\centerline{\bf Acknowledgements}
\bs
 M.C.B. acknowledges the partial support of Funda\c c\~ao para a 
Ci\^encia e a Tecnologia (FCT) under the grant POCTI/1999/FIS/36285. The 
work of A.A. Sen is fully financed by the same grant. N.M.C. Santos is 
supported by FCT grant SFRH/BD/4797/2001.


\end{document}